\documentclass[aps,prd,amsmath,amssymb,preprintnumbers,nofootinbib,a4paper,11pt]{revtex4}
\pdfoutput=1
\usepackage{amsthm}
\usepackage{graphicx}
	\graphicspath{{./NewFig/}}
\usepackage{url}
\usepackage[bookmarks, pagebackref=false]{hyperref}
\usepackage{color}
	\definecolor{rossoCP3}{cmyk}{0,.88,.77,.40}
		\definecolor{graa}{rgb}{0.8,0.8,0.8}
		\definecolor{blaa}{rgb}{0.2,0.2,0.6}
		\hypersetup{
			colorlinks, 
			bookmarksopen, 
			bookmarksnumbered,
			citecolor=blaa, 		
			linkcolor=rossoCP3,	
			urlcolor=rossoCP3,			
			}
\usepackage{dcolumn}
\usepackage{bm}
\usepackage{bbm}
\usepackage{pxfonts}

\usepackage{epsfig}
\usepackage{placeins}

\usepackage{bbold}

\usepackage[ margin=5pt, font=small,labelfont=bf, justification=raggedright]{caption}
\usepackage{youngtab}
\usepackage{slashed}
\usepackage{subfig}

\newcommand{\beq}{\begin{eqnarray}}
\newcommand{\eeq}{\end{eqnarray}}

\newcommand{\bmp}{\noindent\begin{minipage}{16cm}}
\newcommand{\emp}{\end{minipage}\vskip 7mm} 

\def\lsim{\mathrel{\rlap{\lower4pt\hbox{\hskip1pt$\sim$}}
    \raise1pt\hbox{$<$}}}                
\def\gsim{\mathrel{\rlap{\lower4pt\hbox{\hskip1pt$\sim$}}
    \raise1pt\hbox{$>$}}}                

\begin{document}

\title{\LARGE \color{rossoCP3} (Pseudo)Real Vacua}
 \author{Thomas A. Ryttov}\email{ryttov@cp3.dias.sdu.dk} 
  \affiliation{
{ \color{rossoCP3}  \rm CP}$^{\color{rossoCP3} \bf 3}${\color{rossoCP3}\rm-Origins} \& the Danish Institute for Advanced Study {\color{rossoCP3} \rm DIAS},\\ 
University of Southern Denmark, Campusvej 55, DK-5230 Odense M, Denmark.
}

\begin{abstract}
We study the alignment of the vacuum in fermionic gauge theories with enhanced global symmetries by adding a small explicit mass term. Specifically we center our investigations around the theories with two Dirac fermions in a real or pseudoreal representation of the gauge group. We find that as we vary the explicit fermion masses $m_U$ and $m_D$ there are critical points where the system makes a discontinuous jump from one vacuum to another. This occurs when one of the fermion masses vanish. Hence there are four unique vacua depending only on the sign of $m_U$ and $m_D$. Due to the enhanced global symmetry nothing is therefore expected to happen at the point $m_U=-m_D$ where, on the other hand, theories with fermions in a complex representation of the gauge group are expected to develop a CP violating (Dashen) phase.
 \vskip .1cm
{\footnotesize  \it Preprint:  CP$^3$-Origins-2014-028  DNRF 90\ \& DIAS-2014-28}
 \end{abstract}

\maketitle

\newpage
     
\section{Introduction}

Even though theories of strong interactions have been with us for about forty years there are still many outstanding questions that need to be answered. For instance we still lack a complete understanding of such phenomena as confinement, chiral symmetry breaking and infrared fixed points with associated conformal behavior. Solving strongly interacting theories would not only yield a complete understanding of Quantum Chromo Dynamics (QCD) but would also provide invaluable assistance to the search for possible physics beyond the Standard Model. For a review on these aspects see for instance \cite{Sannino:2009za,Hill:2002ap}.

Strong interactions also have the potential to exhibit another interesting dynamical phenomena as first noted by Dashen \cite{Dashen}. It is quite amusing to note that even before the discovery of QCD he had observed that under special circumstances the combination of charge conjugation and parity (CP) could potentially be spontaneously broken \cite{Dashen}. It was discovered that by appropriate choices of the masses of the quarks the vacuum could align in a direction in which CP was broken. Specifically for three quark flavors if the strange quark mass become sufficiently negative the vacuum will pick up nontrivial complex phases and CP will be broken. A CP violating phase can also materialize for only two quark flavors as emphasized by M. Creutz and collaborators \cite{Creutz:1995wf,Creutz:2003xu,Creutz:2003xc,Creutz:2010ts,Creutz:2013xfa,Aoki:2014moa,Creutz:2000bs}. Here it is found that once the mass of the down quark become equal in magnitude but opposite in sign to the mass of the up quark a vacuum degeneracy appears signaling that the Nambu-Goldstone bosons have become massless. At this point it has been suggested that the neutral Nambu-Goldstone boson (the neutral pion), being odd under CP, might condense and spontaneously break CP  \cite{Creutz:1995wf,Creutz:2003xu,Creutz:2003xc,Creutz:2010ts,Creutz:2013xfa,Aoki:2014moa,Creutz:2000bs}. For both two and three quark flavors the  spontaneous CP violation is a consequence of the alignment of the vacuum by an (external) perturbation, i.e. the mass term.

It is only natural to ask whether this persists if we slowly change the various parameters of QCD. First, since vector-like gauge theories with a set of massless fermions in an arbitrary complex representation of the gauge group have the same global chiral symmetry all such systems must share the phenomena of spontaneous CP violation in a manner identical to QCD. Second, it was recently shown that in the four flavor case there is a region similar to the three flavor case in which CP is violated \cite{Ryttov:2014kua}. In fact for any number of  flavors $N_f$ one can show that for appropriately tuned external mass perturbations CP will break spontaneously \cite{Ryttov:2014kua}.

In this work we intend to change yet another set of parameters and search for CP violation. Instead of assuming that the fermions belong to a complex representation of the gauge group we shall turn our attention to two Dirac flavors in a \emph{real} and \emph{pseudoreal} representation. This has the immediate effect of enhancing the global symmetry of the theory and therefore it could potentially lead to a different dynamical behavior. Loosely speaking we will find that once we add a mass perturbation for the two Dirac fermions and drive the vacuum in different directions the system always has the freedom (due to the \emph{enhanced} global symmetry) to choose a vacuum in which CP is preserved. The behavior is quite novel since the system makes a discontinuous jump from one vacuum to another as we dial the masses from positive/negative to negative/positive values. In other words if we try to force the system into a CP violating phase it responds by jumping into a vacuum in which CP is preserved and hence where a fewer number of symmetries are broken.

Examples of theories that fall under our general considerations are $SU(2)$ gauge theories with $N_f=2$ massless fermions in either the fundamental (pseudoreal) representation or the three dimensional adjoint (real) representation. Besides being playgrounds for studying modifications of QCD these theories are also the basis for a great number of attempts to build realistic technicolor models able to explain electroweak symmetry breaking in a natural way \cite{Weinberg:1975gm,Weinberg:1979bn,Sannino:2004qp,Dietrich:2005jn,Dietrich:2005wk,Foadi:2007ue,Ryttov:2008xe,Belyaev:2008yj,Dietrich:2009af,Dietrich:2009ix,Dietrich:2008ni,Gudnason:2006mk}. This is yet another strong motivation for our work.

Situations where the alignment of the vacuum is produced by other perturbations than mass perturbations have also been investigated in the context of technicolor theories. Specifically the question as to how the gauging of the electroweak symmetry, which explicitly breaks the global symmetry, aligns the vacuum \cite{Preskill:1980mz}. In addition some years ago it was suggested that the observed CP violation in the Standard Model could be accounted for in (extended) technicolor theories \cite{Eichten:1980du,Lane:2002wv,Martin:2004ec,Lane:2005we,Lane:2005vp,Lane:2000es,Rador:2009sy}. Here quarks and techniquarks are coupled via four fermion interactions and in this way alignment in the techniquark sector would determine the alignment in the quark sector. 

The paper is organized as follows: In Section \ref{sec:VA} we discuss vacuum alignment in general terms. We also discuss the discrete symmetries C and P for fermions in a real or pseudoreal representation of the gauge group. In Section \ref{sec:real} and \ref{sec:pseudoreal} we present our results for the theory with two Dirac fermions in a real representation and pseudoreal of the gauge group. We then conclude in Section \ref{sec:conclusion}.

\section{Symmetries and Vacuum Alignment}\label{sec:VA}

As our starting point we consider vector-like gauge theories with a set of massless fermions belonging to some irreducible representation of the gauge group. The specific situation where the gauge group is $SU(3)$ and the representation is the fundamental representation corresponds to massless QCD with some number of flavors. 

In general such gauge theories with a fermionic matter sector have a global continuous symmetry that depends on the representation to which the fermions belong. If the representation is complex the global symmetry is $G=SU(N_f)_L \times SU(N_f)_R\times U(1)_B$ and if the representation is real or pseudoreal the global symmetry is enhanced to $G=SU(2N_f)$ \cite{Peskin:1980gc}. Here $N_f$ counts the number of Dirac fermions $\Psi_D = (\psi_L, \psi_R)^T$ where $\psi_L$ and $\psi_R$ are respectively a left-handed and right-handed Weyl fermion. The abelian $U(1)_B$ is the standard baryon number and is anomaly free. 

The reason for the appearance of the enhanced symmetry when the representation is real or pseudoreal is due to the fact that one can always write a right-handed Weyl fermion in a given representation as a left-handed Weyl fermion in the conjugated representation. Then if the representation is real or pseudoreal the conjugated representation is equivalent to the representation itself. Hence if we have $N_f$ Dirac fermions we can consider them as $2N_f$ left-handed Weyl fermions $\lambda_L^f,\ f=1,\ldots, 2N_f$ all in the same real or pseudoreal representation. From now on we shall omit the subscript $L$ and always take the Weyl fermions $\lambda^f$ to be left-handed. 

Gauge theories may also posses the discrete symmetries charge conjugation C, parity P and time reversal T or any combination hereof. Charge conjugation exchanges a particle with its antiparticle or vice versa while parity changes the chirality of a particle. For Dirac fermions in a complex representation of the gauge group these discrete symmetries are implemented in standard fashion and can be found for instance in the appendix of \cite{Ryttov:2014kua}.

For a set of fermions $\lambda^f$ belonging to a real or pseudoreal representation of the gauge group one might first expect them to be Majorana, i.e. that they are equal to their own antiparticles. This follows from the fact that charge conjugation involves a complex conjugation of the fermion fields and the reality or pseudoreality of the fermion representation. However they also transform as a fundamental under global $SU(2N_f)$ and for $N_f>1$ this is a complex representation. Hence under charge conjugation, in which complex conjugation occurs, these fermions cannot be transformed into them selves. Instead under charge conjugation and parity we must consider them in the Dirac basis in which the implementation of all the discrete symmetries are known.

Next consider theories that are asymptotically free and assume that as the energy scale is lowered the gauge interactions become strong such that a bilinear condensate is formed spontaneously breaking the continuous global symmetry $G$ to some subgroup $G'$. Even though the Hamiltonian $\mathcal{H}$ of the system is invariant under all transformations of $G$ the vacuum is only invariant under transformations of $G'$. There is a degeneracy of vacua all parameterized by the elements that belong to $G$ but not to $G'$. As is well known this degeneracy is manifested in the appearance of a set of Nambu-Goldstone bosons.

If we add a small perturbation $\mathcal{H}'$ to the system explicitly breaking the global symmetry $G$ the vacuum degeneracy will be lifted and the vacuum will become \emph{aligned} with the explicit symmetry breaking perturbation. In order to find the aligned (correct) vacuum that is left over we must minimize the total energy $\langle \Omega |g^{\dagger} \left( \mathcal{H} + \mathcal{H}'  \right) g| \Omega \rangle $ over all vacua $g|\Omega \rangle$ where $g\in G/G'$ and $| \Omega \rangle$ is some standard reference vacuum. It should be clear that the contribution $\langle \Omega | g^{\dagger} \mathcal{H} g | \Omega \rangle = \langle \Omega | \mathcal{H} | \Omega \rangle$ is just a constant since the unperturbed Hamiltonian is symmetric under all transformations. Hence we can simplify the procedure of finding the aligned vacuum by considering the minimization of only the perturbation
\begin{eqnarray}
E &=& \langle \Omega | g^{\dagger} \mathcal{H}' g | \Omega \rangle
\end{eqnarray}
It should be noted that we can view the minimization in two different ways: Either we find the rotated vacuum $g | \Omega \rangle$ that minimizes the energy or we find the rotated Hamiltonian $g^{\dagger} \mathcal{H}' g$ that is compatible with the reference vacuum $| \Omega \rangle$. The two views are equivalent.

\section{Real Representations}\label{sec:real}

As one of our main examples we will consider two massless Dirac fermions denoted as  $(U_L,U_R)^T$ (up) and $(D_L,D_R)^T$ (down) transforming according to a real representation of the gauge group. From the above discussion this is equivalent to four Weyl fermions all transforming according to the same real representation
\begin{eqnarray}\label{eq:lambda}
\lambda &=& \left(
\begin{array}{c}
U_L \\
D_L \\
- i \sigma^2 U_R^* \\
- i \sigma^2 D_R^*
\end{array}
\right)
\end{eqnarray}
The Weyl fermions $\lambda^f,\ f=1,\dots,4$ are all treated as left-handed fields. We have also suppressed the gauge index. The theory enjoys an $SU(4)$ global symmetry in which the four Weyl fields are rotated among each other. Assume now that the gauge interactions become strong such that the following condensate is formed \cite{Holdom:1984sk,Yamawaki:1985zg,Appelquist:1986an,Appelquist:1986tr,Appelquist:1987fc,Appelquist:1988yc,Appelquist:1997dc,Brodsky:2008be,Ryttov:2007sr,Ryttov:2007cx,Ryttov:2009yw,Ryttov:2010hs,Ryttov:2010iz,Ryttov:2012qu,Ryttov:2013hka,Ryttov:2013ura,Mojaza:2012zd,Pica:2010mt,Pica:2010xq}
\begin{eqnarray}
\langle \Omega | \lambda^f \lambda^{f'} | \Omega \rangle &=& - \frac{1}{2} \Delta E^{ff'} \ , \qquad E=
\left(
\begin{array}{cccc}
0 & 0 & 1 & 0 \\
0 & 0 & 0 & 1 \\
1 & 0 & 0 & 0 \\
0 & 1 & 0 & 0 
\end{array}
\right)
\end{eqnarray}
where for brevity we have defined $\lambda \lambda = - \lambda^T i \sigma^2 \lambda$. In terms of the up and down fermion fields we can also write the condensate in the standard form $\langle \Omega | \lambda^f \lambda^{f'} | \Omega \rangle E_{ff'}= 2 \langle \Omega | \overline{U}_R U_L + \overline{D}_R D_L   | \Omega \rangle $ by contracting the flavor indices with $E_{ff'}$. This justifies our choice of the specific reference vacuum $E$ since here the condensate is particularly simple. Also the notation is such that $\overline{U}_R = U_R^*$ and $\overline{D}_R=D_R^*$. 

Since the condensate is symmetric in its indices it is invariant only under $SO(4)$ transformations and therefore spontaneously breaks the $SU(4)$ global symmetry. This spontaneous symmetry breaking leads to the well known appearance of $15-6=9$ Nambu-Goldstone bosons. 

Let us try to force the vacuum in a different direction by adding a small mass term. Such a perturbation explicitly breaks the $SU(4)$ global symmetry, lifts the vacuum degeneracy and provides the Nambu-Goldstone bosons with a mass. To be concrete assume that a Dirac mass term is added to the theory
\begin{eqnarray}
\mathcal{H}' &=& \frac{1}{2} \lambda M^{\dagger} \lambda + \text{h.c.} \ , \qquad \qquad M = M^{\dagger} = 
\left( 
\begin{array}{cccc}
0 & 0 & m_U & 0 \\
0 & 0 & 0 & m_D \\
m_U & 0 & 0 & 0 \\
0 & m_D & 0 & 0 
\end{array}
\right)
\end{eqnarray}
Here the Dirac mass matrix is symmetric and each entry $m_U$ and $m_D$ are taken to be real such that it preserves CP. In order to find the correct aligned vacuum we need to follow the above steps of minimizing the energy
\begin{eqnarray}
E(V) &=& \langle \Omega | \frac{1}{2}(g \lambda ) M (g\lambda) | \Omega \rangle + \text{h.c.} = - \frac{\Delta}{4}  \text{Tr} \left[ M^{\dagger} V+MV^{\dagger}  \right]
\end{eqnarray}
where the rotated vacuum is $V=gEg^T$ and $g\in G/G'$. We must find the rotated (aligned) vacuum $V$ that minimizes the energy. To this end we first note that $V$ must be of the same form as the Dirac mass matrix since the energy involves a trace. Second we note that the vacuum must be symmetric since the reference vacuum $E$ is symmetric. In addition it must be unitary and can therefore only depend on two phases $e^{i\theta_U}$ and $e^{i \theta_D}$. Lastly since the vacuum must have unit determinant the two angles must up to a multiple of $\pi$. Putting it all together we find that the aligned vacuum must be of the form
\begin{eqnarray}
V &=& 
\left( 
\begin{array}{cccc}
0 & 0 & e^{i \theta_U} & 0 \\
0 & 0 & 0 & e^{i \theta_D} \\
e^{i \theta_U} & 0 & 0 & 0 \\
0 & e^{i\theta_D} & 0 & 0
\end{array}
\right) \ , \qquad \theta_U + \theta_D = \pi n
\end{eqnarray}
where $n$ is some integer number. It is important to observe that the angles should add up to \emph{any} multiple of $\pi$. This is in contrast to the case where the fermions belong to a complex representation of the gauge group \cite{Creutz:1995wf,Creutz:2003xu,Creutz:2003xc,Creutz:2010ts,Creutz:2013xfa,Aoki:2014moa,Creutz:2000bs,Ryttov:2014kua}. Here the vacuum angles must add up to an \emph{even} multiple of $\pi$. As we shall see below this has a major impact on the behavior of the system as we vary the fermion masses aligning the vacuum in different directions. This differece is a direct consequence of the enhanced global symmetry. 

It should be clear that without loss of generality we can take $n=0$ or $n=1$. In this way we can speak about two different branches of the system in which the energy is
\begin{eqnarray}
E(V) &=& \left\{
\begin{array}{l}
-\Delta \left( m_U + m_D \right) \cos \theta_U \ , \qquad \theta_D = - \theta_U \ , \qquad \quad n=0 \\
- \Delta \left( m_U - m_D  \right) \cos \theta_U \ , \qquad \theta_D = \pi - \theta_U \ , \qquad n=1 
\end{array}
\right.
\end{eqnarray}
We want to minimize the energy over the vacuum angle $\theta_U$. The location of the minima depends on the sign of the up and down fermion masses. If both fermion masses have the same sign the minima are located on the $n=0$ branch. If they are both positive they are found at $\theta_U$ being equal to an even multiple of $\pi$ and if they are both negative they are found at $\theta_U$ being equal to an odd multiple of $\pi$. On the other hand the fermion masses might also differ in sign. In that case the minima are located on the $n=1$ branch. If the up fermion mass is positive and the down fermion mass is negative then the minima are found at $\theta_U$ being equal to an even multiple of $\pi$ while in the opposite case where the up fermion mass is negative and the down fermion mass is positive the minima are found at $\theta_U$ being equal to an odd multiple of $\pi$. Summarizing we therefore find that the aligned vacuum is
\begin{eqnarray}
V_{++} &=& \left( 
\begin{array}{cccc}
0 & 0 & 1 & 0 \\
0 & 0 & 0 & 1 \\
1 & 0 & 0 & 0 \\
0 & 1 & 0 & 0 
\end{array}
\right) \ , \qquad \qquad \quad E(V_{++}) = - \Delta \left( | m_U| + |m_D | \right)  \ , \qquad m_U,m_D > 0 \\
V_{--} &=&  
\left( 
\begin{array}{cccc}
0 & 0 & -1 & 0 \\
0 & 0 & 0 & -1 \\
-1 & 0 & 0 & 0 \\
0 & -1 & 0 & 0
\end{array}
\right) \ , \qquad \ E(V_{--}) =  - \Delta \left( | m_U| + |m_D | \right)  \ , \qquad m_U,m_D < 0 \\
V_{+-} &=& \left( 
\begin{array}{cccc}
0 & 0 & 1 & 0  \\
0 & 0 & 0 & -1\\
1 & 0 & 0 & 0 \\
0 & -1 & 0 & 0
\end{array}
\right) \ , \qquad \qquad E(V_{+-}) = - \Delta \left( |m_U|+|m_D|  \right) \ , \qquad m_U>0,m_D<0 \\
V_{-+} &=& \left( 
\begin{array}{cccc}
0 & 0 & -1 & 0 \\
0 & 0 & 0 & 1 \\
-1 & 0 & 0 & 0 \\
0 & 1 & 0 & 0  
\end{array}
\right) \ , \qquad \qquad E(V_{--}) = - \Delta \left( |m_U|+|m_D| \right) \ , \qquad m_U<0,m_D>0 
\end{eqnarray}

We are now in a position to study the response of the system to the external mass perturbation. First imagine that both up and down explicit fermion mass terms $m_U$ and $m_D$ are positive. Then the system must be found to reside in the single unique vacuum $V_{++}$ which is on the $n=0$ branch. This is for arbitrary positive values $m_U,m_D>0$. Let us now fix $m_U$ and slowly dial the down fermion mass $m_D$ towards zero. As the down fermion mass passes through the critical value $m_D=0$ the system becomes discretely degenerate and a new additional vacuum $V_{+-}$ appears on the $n=1$ branch. If we continue to decrease the down fermion mass to negative values the discrete degeneracy disappears and the vacuum of the system uniquely becomes $V_{+-}$. Nothing is then expected to further happen for any negative value $m_D<0$. 

\begin{figure}[bt]
\centering
\includegraphics[width=0.3\columnwidth]{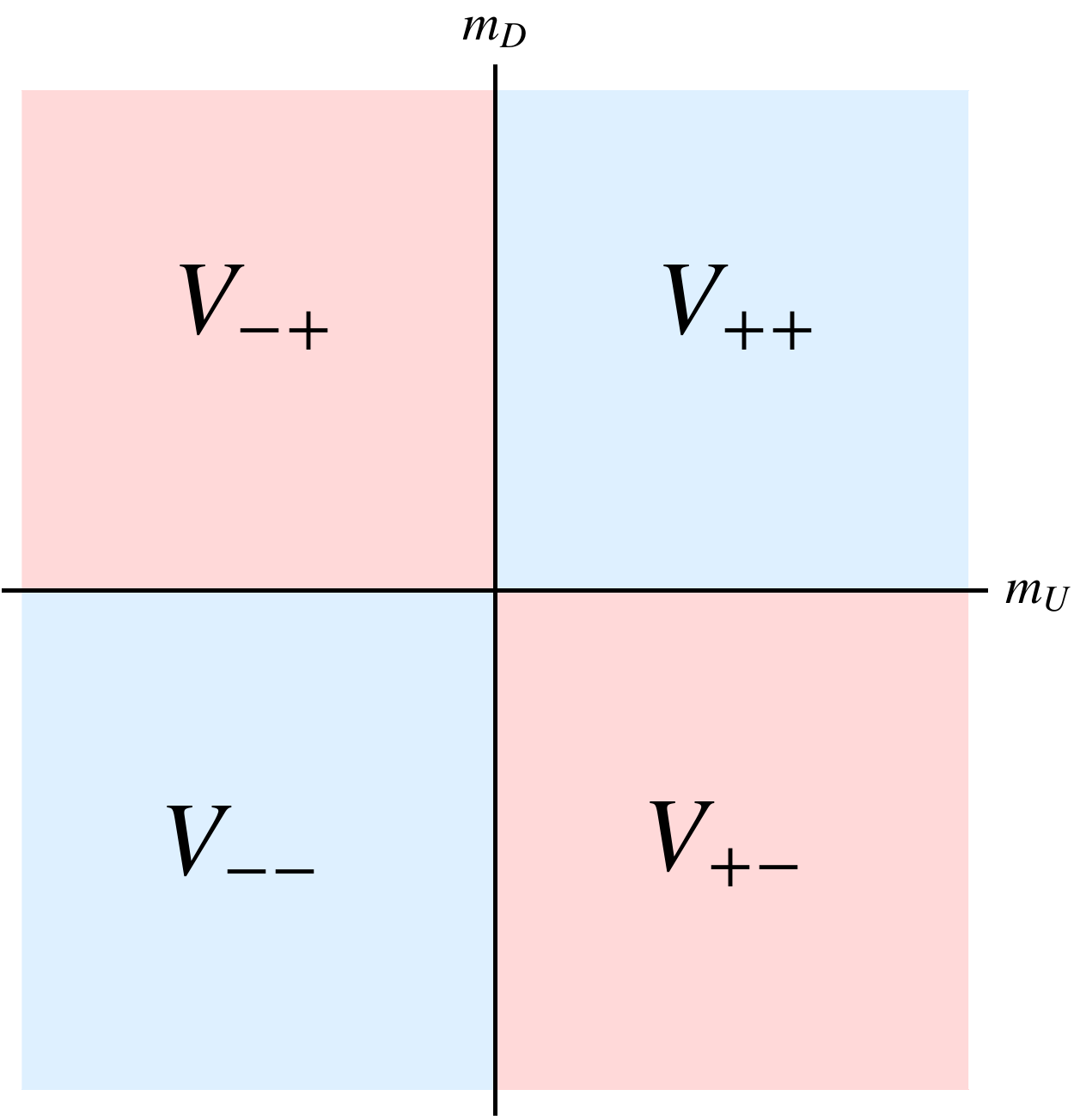}  
 \caption{A plot of the vacuum structure (with $V_{++}=-V_{--}$ and $V_{+-} = -V_{-+}$) as function of the up and down mass $m_U$ and $m_D$ with two Dirac flavors in a real representation of the gauge group. When crossing from one vacuum into another the system makes a discontinuous jump. Blue corresponds to the $n=0$ branch while red corresponds to the $n=1$ branch. }
\label{RealPlot}
\end{figure}

What has happened is quite novel since the system must have gone through a discontinuous jump at the critical value where the down fermion mass vanishes. As we cross the critical value the system must respond by jumping from the $n=0$ branch to the $n=1$ branch. It should be clear that the above discontinuous behavior occurs for an arbitrary magnitude and sign of $m_U$ and $m_D$. It does not matter whether $m_U$ and $m_D$ initially are positive or negative. If we fix one of them and dial the other then as the second mass passes through zero the system must perform a discontinuous jump from the $n=0$ ($n=1$) branch to the $n=1$ ($n=0$) branch. In Fig. \ref{RealPlot} we plot the vacuum structure in terms of the up and down explicit fermion mass.

Irrespective of which of the four vacua the system resides in the continuous global symmetry $SU(4)$ is broken to $SO(4)$.\footnote{The fact that these four vacua preserve an $SO(4)$ symmetry can be checked by using the generators found \cite{Foadi:2007ue}. } The embedding (or direction) of the $SO(4)$ subgroup of course depends on the specific vacuum. Note also that in all four vacua there are no nontrivial complex phases. In this way discrete CP symmetry is preserved for arbitrary mass perturbations, i.e. we cannot drive the vacuum into a direction in which CP is spontaneously broken. 

Let us contrast this to the case where the fermions are in a complex representation of the gauge group and for which the continuous global symmetry is $SU(N_f)_L \times SU(N_f)_R \times U(1)_B$. We will discuss the two flavor case since this is the situation that mimics our example above \cite{Dashen,Creutz:1995wf,Creutz:2003xu,Creutz:2003xc,Creutz:2010ts,Creutz:2013xfa,Aoki:2014moa,Creutz:2000bs}. However as we shall see there are crucial differences. 

Consider therefore two Dirac flavors, up and down, in a complex representation of the gauge group and take their mass perturbations to be positive. The vacuum is then aligned in the direction $\text{diag}(1,1)$. But if we fix the up fermion mass and slowly dial the down fermion mass towards zero no new vacua appear \cite{Dashen,Creutz:1995wf,Creutz:2003xu,Creutz:2003xc,Creutz:2010ts,Creutz:2013xfa,Aoki:2014moa,Creutz:2000bs}. One might naively have expected that the system would choose the vacuum $\text{diag}(1,-1)$ but this does not have unit determinant so it is not a possibility. Instead the vacuum continues to be $\text{diag}(1,1)$ until the critical point where the up and down fermion masses are equal in magnitude but opposite in sign is encountered. The vacuum is here forced to take complex values $\text{diag}(e^{i\theta},e^{-i\theta})$ where $\theta$ is some arbitrary angle. Again it cannot be driven in the direction $\text{diag}(1,-1)$ since it does not have unit determinant. Instead a continuous vacuum degeneracy appears indicating that the Nambu-Goldstone bosons have once again become massless. At this critical point it has been suggested that the neutral Nambu-Goldstone boson (the neutral pion), being odd under CP, develops a VEV spontaneously breaking CP \cite{Dashen,Creutz:1995wf,Creutz:2003xu,Creutz:2003xc,Creutz:2010ts,Creutz:2013xfa,Aoki:2014moa,Creutz:2000bs}. This is the Dashen phase. For completeness we note that as the down fermion mass passes the critical point and decreases even further the vacuum is aligned in the direction $\text{diag}(-1,-1)$ and the vacuum degeneracy disappears.

This differs in essential aspects from the case where the fermions belong to a real representation of the gauge group and the system has an enhanced global symmetry. Here we find no evidence for the existence of a Dashen type phase. Technically it is due to the fact that the system has two different branches $n=0$ and $n=1$ and that as we dial the fermion masses the system can undergo discontinuous jumps from one branch to another. If we for the moment imagine that the system did not have the $n=1$ branch then the behavior would be just as in the case where the fermion representation is complex. As we pass the down fermion mass through zero nothing is expected to happen. Then again at the point where the down and up fermion masses are equal in magnitude but opposite in sign a vacuum degeneracy appears since the energy of the system vanishes identically for any value of the vacuum angles. Here the Nambu-Goldstone bosons become massless and via a mechanism similar to the case for complex representations where the neutral one develops a VEV could break CP spontaneously. However this is not observed.

\section{Pseudoreal Representations}\label{sec:pseudoreal}

In this section we complete our study by considering a similar setup as above but with fermions belonging to a pseudoreal representation of the gauge group. We again take two Dirac flavors and arrange them into a set of left-handed transforming fermions $\lambda^f,\ f=1,\ldots,4$. The theory enjoys a continuous global symmetry $G=SU(4)$ and a set of discrete symmetries. As above we imagine that at some energy scale the gauge interactions become strong and the following condensate is formed
\begin{eqnarray}
\langle \Omega | \lambda^f \lambda^{f'} | \Omega \rangle &=& - \frac{1}{2} \Delta E^{ff'} \ , \qquad E= 
\left(
\begin{array}{cccc}
0 & 0 & 1 & 0 \\
0 & 0 & 0 & 1 \\
-1 & 0 & 0 & 0 \\
0 & -1 & 0 & 0
\end{array}
\right)
\end{eqnarray}
Since the condensate is antisymmetric in its indices it leaves invariant only a $G'=Sp(4)$ subgroup. Hence it breaks the continuous global symmetry spontaneously and is the source for the appearance of $15-10=5$ Nambu-Goldstone bosons. Note again that if we write $\lambda^f$ in terms of the two Dirac fermions as in Eq. \ref{eq:lambda} the condensate can be written as $\langle \Omega | \lambda^f \lambda^{f'} | \Omega \rangle E_{ff'}= 2 \langle \Omega | \overline{U}_R U_L + \overline{D}_R D_L | \Omega \rangle$ justifying our choice of the reference vacuum $E$ since the condensate is here particularly simple. Adding to the theory the following explicit Dirac mass term
\begin{eqnarray}
\mathcal{H}' &=&\frac{1}{2} \lambda M^{\dagger} \lambda  + \text{h.c.} \ , \qquad M=-M^{\dagger} = 
\left(
\begin{array}{cccc}
0 & 0 & m_U &  0 \\
0 & 0 & 0 & m_D \\
-m_U & 0 & 0 & 0 \\
0 & -m_D & 0 & 0 
\end{array}
\right)
\end{eqnarray}
we want to find the rotated vacuum $V=g E g^T,\ g \in G/G'$ that minimizes the energy
\begin{eqnarray}
E(V) &=& \langle \Omega | \frac{1}{2} (g \lambda ) M^{\dagger} (g \lambda) | \Omega \rangle + \text{h.c.} = - \frac{\Delta}{4} \text{Tr} \left[ M^{\dagger} V + M V^{\dagger} \right] 
\end{eqnarray}
Both the up and down fermion mass terms $m_U$ and $m_D$ are taken to be real. Via a similar line of arguments as in the case with a real fermion representation we find that the rotated vacuum depends on two angles $\theta_U$ and $\theta_D$ that adds to a multiple of $\pi$ and be of the form
\begin{eqnarray}
V &=& \left(
\begin{array}{cccc}
0 & 0 & e^{i \theta_U} & 0 \\
0 & 0 & 0 & e^{i \theta_D} \\
- e^{i\theta_U} & 0 & 0 & 0 \\
0 & - e^{i \theta_D} & 0 & 0
\end{array}
\right) \ , \qquad \theta_U + \theta_D = \pi n
\end{eqnarray}
Again there are two different branches corresponding to $n=0$ and $n=1$. All other branches are equivalent to one of these. Hence we want to minimize
\begin{eqnarray}
E(V) &=& \left\{
\begin{array}{l}
- \Delta \left( m_U + m_D \right) \cos \theta_U \ , \qquad \theta_D = -\theta_U \ , \qquad\quad n=0 \\
- \Delta \left( m_U - m_D \right) \cos \theta_U \ , \qquad \theta_D = \pi -\theta_U \ , \qquad n=1 
\end{array}
\right.
\end{eqnarray}
over the vacuum angle $\theta_U$. Depending on the sign of the up and down fermion masses the minima are located at the same values of the vacuum angles as in the case with a real fermion representation
\begin{eqnarray}
V_{++} &=& \left(
\begin{array}{cccc}
0 & 0 & 1 & 0  \\
0 & 0 & 0 & 1\\
-1 & 0 & 0 & 0 \\
0 & -1 & 0 & 0
\end{array}
\right) \ , \qquad  E(V_{++})= - \Delta \left( |m_U|+|m_D| \right)\ , \qquad m_U,m_D>0  \\
V_{--} &=& \left(
\begin{array}{cccc}
0 & 0 & -1 & 0  \\
0 & 0 & 0 & -1\\
1 & 0 & 0 & 0 \\
0 & 1 & 0 & 0
\end{array}
\right) \ , \qquad  E(V_{--})= - \Delta \left( |m_U|+|m_D| \right)\ , \qquad m_U,m_D<0 \\
V_{+-} &=& \left(
\begin{array}{cccc}
0 & 0 & 1 & 0  \\
0 & 0 & 0 & -1\\
-1 & 0 & 0 & 0 \\
0 & 1 & 0 & 0
\end{array}
\right) \ , \qquad  E(V_{+-})= - \Delta \left( |m_U|+|m_D| \right)\ , \qquad m_U>0,m_D<0 \\
V_{-+} &=& \left(
\begin{array}{cccc}
0 & 0 & -1 & 0  \\
0 & 0 & 0 & 1\\
1 & 0 & 0 & 0 \\
0 & -1 & 0 & 0
\end{array}
\right) \ , \qquad  E(V_{-+})= - \Delta \left( |m_U|+|m_D| \right)\ , \qquad m_U<0,m_D>0 
\end{eqnarray}
The behavior is identical to the case of a real fermion representation. If we fix the mass of one of the fermions and dial the other one then at the critical point where it vanishes the system makes a discontinuous jump from one branch to the other. There is no critical point where all the Nambu-Goldstone bosons become massless with a vacuum degeneracy reappearing. Hence it is not possible with a Dashen type phase where CP is spontaneously broken when the fermions belong to a pseudoreal representation of the gauge group.

\section{Conclusion}\label{sec:conclusion}

We investigated the vacuum structure of fermionic gauge theories with two Dirac favors by adding a small explicit mass perturbation. We picked the fermions to transform according to a real or pseudoreal representation of the gauge group. By dialing the explicit masses we studied how the vacuum aligned in different directions. We were motivated by the fact that when the fermions are in a complex representation of the gauge group the system has the potential to develop a so called Dashen phase where CP is spontaneously broken. We found that this is not an option if one considers theories with enhanced global symmetries. In some sense the global symmetry is so large that if one attempts to force the system in a CP violating direction it has the freedom to align in a direction in which CP is preserved. In other words the system prefers to break the least number of symmetries.

\acknowledgments
The author would like to thank F. Sannino for discussions and careful reading of the manuscript. The author would also like to thank the CERN Theory Division for kind hospitality while this work was completed. The CP$^3$-Origins centre is partially funded by the Danish National Research Foundation, grant number DNRF90.

\end{document}